\documentclass[fleqn,twoside,psfig]{article}
\usepackage{espcrc2}

\usepackage{psfig}
\newcommand{\beq}{\begin{equation}}
\newcommand{\eeq}{\end{equation}}
\newcommand{\beqn}{\begin{eqnarray}}
\newcommand{\eeqn}{\end{eqnarray}}
\newcommand{\beqns}{\begin{eqnarray*}}
\newcommand{\eeqns}{\end{eqnarray*}}

\newcommand{\hm}{\hspace{-0.05cm}}

\newcommand{\intl}{\int\limits}

\newcommand{\e}{\epsilon}
\def\PL{{\it Phys. Lett.}}

\def\PRL{{\it Phys. Rev. Lett.}}

\def\ea{{\it et al.}}
\def\Cl{Collaboration}
\def\pc{$\%$}

\def\sfs{spectral functions}

\def\as{$\alpha_s$}
\def\asm{$\alpha_s(m_\tau^2)$}

\def\ee{$e^+e^-$}

\def\FOPTCI{$\rm FOPT_{\rm CI}$}
\def\Rt{$R_\tau$}

\def\RtVA{$R_{\tau,V/A}$}

\def\RtVpA{$R_{\tau,V+A}$}

\def\Rts{$R_\tau(s_0)$}

\def\RtVpAs{$R_{\tau,V+A}(s_0)$}

\def\piz{\pi^0 }
\def\ie{{\it i.e.}}

\title{Spectral Functions from Hadronic $\tau$ Decays and QCD}

\author{Michel Davier\\
	Laboratoire de l'Acc\'el\'erateur Lin\'eaire\\
        IN2P3/CNRS et Universit\'e de Paris-Sud\\
        91898 Orsay, France\\
	E-mail: davier@lal.in2p3.fr}

\begin{document}

\begin{abstract}
Hadronic decays of the $\tau$ lepton provide a clean environment to
study hadron dynamics in an energy regime dominated by resonances, with the
interesting information captured in the spectral functions. Recent results 
from ALEPH on exclusive channels are presented, with some emphasis 
on the $\pi \piz$ final state which plays a crucial role for the determination 
of the hadronic contribution to the muon anomalous magnetic moment. 
Inclusive spectral functions are the basis for QCD analyses, 
delivering an accurate determination of the strong
coupling constant, quantitative information on nonperturbative
contributions and a measurement of the mass of the strange quark.
\vspace{1pc}
\end{abstract}

\maketitle

\section{Introduction}

Hadrons produced in $\tau$ decays are born out of the charged weak
current, {\it i.e.} out of the QCD vacuum. This property guarantees
that hadronic physics factorizes in these processes which are then
completely characterized for each decay channel
by spectral functions as far as the total
rate is concerned . Furthermore, the produced
hadronic systems have $I=1$ and spin-parity $J^P=0^+,1^-$ (V) and
$J^P=0^-,1^+$ (A). The spectral functions are directly related to the
invariant mass spectra of the hadronic final states, normalized
to their respective branching ratios and corrected for the $\tau$
decay kinematics. For a given spin-1 vector decay, one has
\begin{eqnarray}
\label{eq_sf}
   v(s) 
   &\equiv&
           \frac{m_\tau^2}{6\,|V_{ud}|^2\,S_{\mathrm{EW}}}\,
              \frac{B(\tau^-\rightarrow {V^-}\,\nu_\tau)}
                   {B(\tau^-\rightarrow e^-\,\bar{\nu}_e\nu_\tau)} \nonumber \\
   & & \hspace{-1.2cm}        
              \times\frac{d N_{V}}{N_{V}\,ds}\,
              \left[ \left(1-\frac{s}{m_\tau^2}\right)^{\!\!2}\,
                     \left(1+\frac{2s}{M_\tau^2}\right)
              \right]^{-1}\hspace{-0.3cm}
\end{eqnarray}
where $V_{ud}=0.9748 \pm 0.0010$ denotes the CKM 
weak mixing matrix element~\cite{pdg2002} and $S_{\mathrm{EW}}$ 
accounts for electroweak radiative corrections (see below). 
Isospin symmetry (CVC) connects the $\tau$ and $e^+e^-$ annihilation
spectral functions, the latter being proportional to the R ratio.

Hadronic $\tau$ decays are a clean probe of hadron dynamics in an
interesting energy region dominated by resonances. However, perturbative
QCD can be seriously considered due to the relatively large $\tau$ mass.
Many hadronic modes have been measured and studied, while some earlier
discrepancies (before 1990) have been resolved with high-statistics and
low-systematics experiments. Samples of $\sim 4 \times 10^5$ measured decays
are available in each LEP experiment and CLEO. Conditions for low 
systematic uncertainties are particularly well met at LEP: measured
samples have small non-$\tau$ backgrounds ($\sim 1\%$) and large selection
efficiency ($92\%$), for example in ALEPH.

Recent results in the field are discussed in this report.

\section{New ALEPH Spectral Functions}

\begin{figure}[t]
   \centerline{\psfig{file=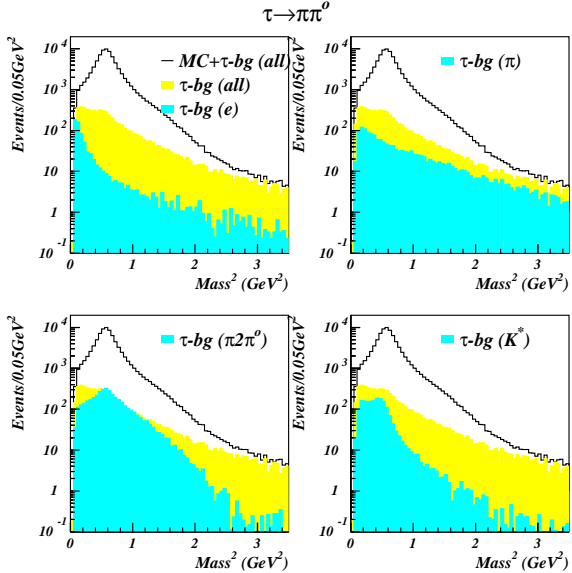,width=75mm}}
   \caption{Feedthrough in the $\pi \pi^0$ channel from other $\tau$ 
decay modes in the ALEPH analysis, as estimated from the Monte Carlo 
simulation. The light-shaded histogram indicates the total feedthrough, 
while the dark-shaded histograms in each plot show the dominant 
individual sources.}
\label{pipi0_back}
\end{figure}

Preliminary spectral functions based on the full LEP1 statistics are 
available from ALEPH. The corresponding results for the branching fractions
are given separately~\cite{tau02_br}. The analysis uses an improved treatment
of photons as compared to the published analyses based on a reduced 
sample~\cite{aleph_v,aleph_a}. Spectral functions are unfolded from the
measured mass spectra after background subtraction using a mass-migration 
matrix obtained from the simulation in order to account for detector 
and reconstruction biases. Backgrounds from non-$\tau \tau$ events 
are small ($< 1 \%$ and subtractions are dominated by $\tau$ decay 
feedthroughs. This is illustrated in Fig.~\ref{pipi0_back} 
for the $\pi \piz$ channel, where the main feedthrough contributions
are from the $\pi$ mode at large masses (extra photon from radiation and
fake photons from hadronic interactions in the electromagnetic calorimeter)
and mostly the $a_1 \rightarrow \pi 2\piz$ mode (one $\piz$ lost).

\begin{figure}[t]
   \centerline{\psfig{file=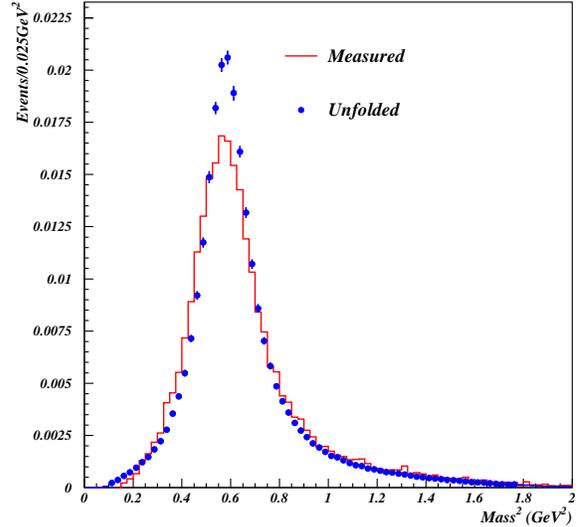,width=75mm}}
   \caption{Spectral functions in the $\pi \pi^0$ mode from ALEPH
before (histogram) and after (points) the unfolding procedure. }
\label{unfold_rho}
\end{figure}

The raw and unfolded spectra for the $\pi^-\pi^0$ mode are given
in Fig.~\ref{unfold_rho}. Apart from the normalization effects studied in
the branching ratio analysis~\cite{tau02_br}, systematic biases affecting
the mass spectra, are studied separately. The corresponding uncertainties 
and correlations in mass are incorporated in covariance matrices for every
considered channel. According to the number of pions in the final state, the
spectral functions are separated into vector and axial-vector components. 
As for final states involving $K\overline{K}$ pairs, specific input 
is required to achieve the $V-A$ separation~\cite{aleph_ksum}. 
Strange final states are treated separately. 

\section{Specific Final States}

\subsection{The $2\pi$ Vector State}

\subsubsection{The Data}

\begin{figure}[t]
   \centerline{\psfig{file=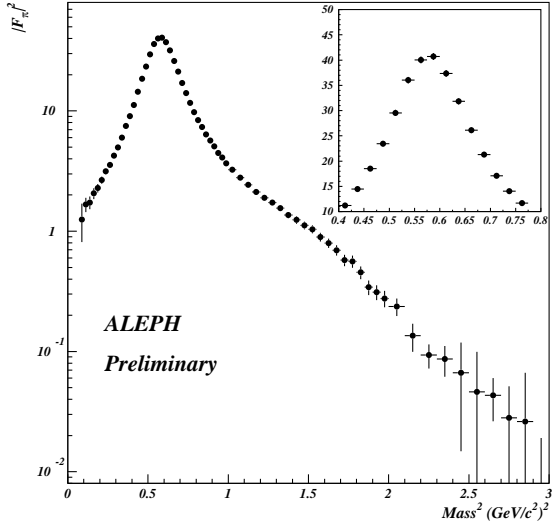,width=80mm}}
   \caption{The pion form factor squared $|F_\pi^-(s)|^2$ measured by ALEPH. 
   Because of the unfolding procedure, the errors in nearby bins are 
   strongly correlated (only diagonal errors are shown).}
\label{aleph_2pi}
\end{figure}

The spectral function from $\tau \rightarrow \nu_\tau \pi^- \pi^0$ 
in the full-LEP1 ALEPH analysis ($\sim 10^5$ events) is presented
in Fig~\ref{aleph_2pi}. It is dominated by the $\rho$(770) resonance,
with some broad structure around 1.25~GeV.

\begin{figure}[t]
   \centerline{\psfig{file=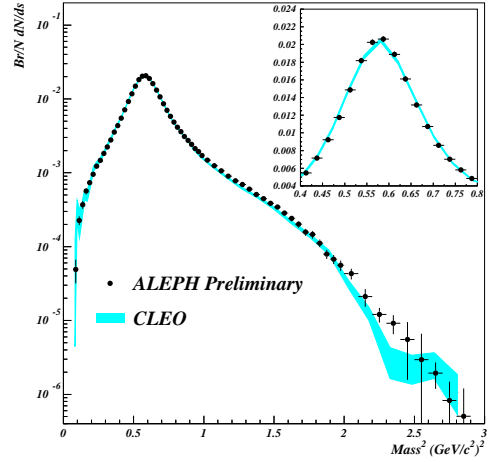,width=70mm}}
   \caption{The comparison between ALEPH and CLEO spectral functions
in the $\pi \pi^0$ mode. }
\label{aleph_cleo_2pi}
\end{figure}

Results from CLEO~\cite{cleo_2pi} are in good agreement with the ALEPH
data. The statistics is comparable in the two cases, however whereas the
acceptance is very flat across the mass spectrum in the case of ALEPH, 
a strong increase is observed for the CLEO case. As a consequence, ALEPH
data are more precise below the $\rho$ peak, while CLEO is more precise
above. Thus the two pieces of data are complementary. 
The comparison is given in Fig.~\ref{aleph_cleo_2pi}.

\subsubsection{$\pi \pi$ Spectral Functions and $\pi$ Form Factors}

It is useful to carefully write down all the factors involved in the 
comparison of \ee\ and $\tau$ spectral functions in order to make
explicit the possible sources of CVC breaking. On the \ee\ side we have
\begin{eqnarray}
\sigma (e^+e^-\longrightarrow \pi^+\pi^-)&=&\frac{4\pi\alpha^2}{s} v_0(s)\\
    v_0(s)&=&\frac {\beta_0^3(s)} {12 \pi} |F^0_\pi(s)|^2 \nonumber
\label{ee_ff}
\end{eqnarray}
where $\beta_0^3(s)$ is the threshold kinematic factor and $F^0_\pi(s)$
the pion form factor. On the $\tau$ side, the physics is contained in the
hadronic mass distribution and one can write correspondingly 
to Eq.~(\ref{ee_ff})
\begin{eqnarray}
\frac {1}{\Gamma} \frac {d \Gamma}{ds}
 (\tau \longrightarrow \pi^-\pi^0 \nu_\tau) &=& \nonumber \\
         & & \hspace{-2.7cm}
                    \frac {6 \pi |V_{ud}|^2 S_{EW}}{m_\tau^2}
                    \frac {B_e}{B_{\pi \pi^0}} C(s) v_-(s) \\
      v_-(s) &=& \frac {\beta_-^3(s)} {12 \pi} |F^-_\pi(s)|^2 \nonumber \\
      C(s) &=& \left(1- \frac {s}{m_\tau^2}\right) \nonumber
       \left(1 + \frac {2 s}{m_\tau^2}\right)
\label{tau_ff}
\end{eqnarray}
SU(2) symmetry (CVC) implies
\beq
v_-(s) =  v_0(s)
\eeq

It is important to note that experiments on $\tau$ decays measure the rate
inclusive of radiative photons, {\it i.e.} for 
$\tau \rightarrow \nu_\tau \pi \piz (\gamma)$. The measured spectral function
is thus $v_-^*(s) =  v_-(s)~G(s)$, where $G(s)$ is a radiative correction.
Also, as $|F^0_\pi(s)|^2$ should only involve the pion hadronic structure, 
it must not contain vacuum polarization in the photon propagator.

Three levels of SU(2) breaking can be identified:
\begin{itemize}
\item {\it electroweak radiative corrections to $\tau$ decays} 
are contained in the $S_{\mathrm{EW}}$ 
factor~\cite{marciano,braaten} which is dominated by short-distance effects.
As such it is expected to be weakly dependent on the specific hadronic
final state, as verified by detailed computations in the 
$\tau \longrightarrow (\pi, K) \nu_\tau$ channels~\cite{decker}. 
Recently, detailed calculations have been performed for the $\pi \pi^0$
channel~\cite{cen} which also confirm the relative smallness of the
long-distance contributions. One can write the total correction as
\beq
 S_{EW} = \frac {S_{EW}^{\rm had} S_{EM}^{\rm had} }{S_{EM}^{\rm lep}}
\eeq
where $S_{EW}^{\rm had}$ is the leading-log short-distance electroweak
factor (which vanishes for leptons) and $S_{EM}^{\rm had,lep}$ are the
nonleading electromagnetic corrections. The latter corrections are
calculated in Ref.~\cite{braaten} at the quark level and in
Ref.~\cite{cen} at the hadron level for the $\pi \pi^0$ decay mode,
and in Ref.~\cite{marciano,braaten} for leptons. The total correction
amounts~\cite{dehz} to $S_{EW}^{\rm inclu} = 1.0198 \pm 0.0006$ 
for the inclusive hadron decay rate and 
$S_{EW}^{\pi \pi^0} = (1.0232 \pm 0.0006)~S_{EM}^{\pi \pi^0}(s)$ 
for the $\pi \pi^0$ decay mode, where $S_{EM}^{\pi \pi^0}(s)$ is an
$s$-dependent radiative correction~\cite{cen}.
\item {\it pion mass splitting}, which is almost completely from
electromagnetic origin, directly breaks isospin symmetry in 
the spectral functions~\cite{adh,czyz} since $\beta_-(s) \neq \beta_0(s)$.
\item symmetry is also broken {\it in the pion form factor}
~\cite{adh,cen}. The $\rho$ width is affected by $\pi$ and 
$\rho$ mass splittings and by explicit electromagnetic decays such as 
$\pi \gamma$, $\eta \gamma$, $l^+l^-$ and $\pi \pi \gamma$.
Isospin violation in the strong amplitude is expected to be negligible
because of the small absolute mass difference between $u$ and $d$ quarks. 
\end{itemize}

\subsubsection{Fitting the $2\pi$ Spectral Function}

The $2\pi$ spectral function is dominated by the $\rho$ resonance. This
provides us with an opportunity to study the relevant description of a
wide hadronic resonance. The Gounaris-Sakurai~\cite{gounaris} (GS)
parametrization satisfies analyticity and unitarity through 
finite width corrections. The line shape has to be modified to account 
for the effect of $\rho-\omega$ interference.

The pion form factor is fitted with interfering amplitudes 
from $\rho$, $\rho'$ and $\rho''$ vector mesons with relative strengths 
1, $\beta$ and $\gamma$. A phase $\phi_\beta$ is also considered, 
since the relative phase of the $\rho$ and $\rho'$ amplitude is a priori 
unknown. In practice we fit the $F_\pi^0(s)$ form factor from 
$e^+e^-$ data and the $F_\pi^-(s)$ form factor
from the $\tau$ spectral function duly corrected
for SU(2) breaking, however only in the spectral function phase space 
and in the $S_{EW}$ factor. In this way, the two form factors can be
readily compared, {\it i.e.} the masses and widths of the dominating 
$\rho$ resonance in the two isospin states can be unambiguously
determined.
On the \ee\ side, all available data have been used 
(complete references can be found in Ref.~\cite{dehz}, 
except those of NA7~\cite{na7} which are 
questionable~\cite{rolandi}. However, in the $\rho$ mass region the new 
precise results from CMD-2~\cite{cmd2} are dominating. On the $\tau$ side, 
the accurate data from ALEPH and CLEO are used.

The systematic uncertainties are included in the fits through appropriate
covariance matrices. The $\rho$ mass systematic uncertainty in $\tau$ data
is mostly from calibration (0.7 MeV for ALEPH and 0.9 MeV for CLEO). The
corresponding uncertainties on the $\rho$ width are 0.8 and 0.7 MeV. Both
mass and width determinations are systematically limited in $\tau$ data,
however they are uncorrelated between ALEPH and CLEO. Due to the large
event statistics, the fits are quite sensitive to the precise line shape
and to the interference between the different amplitudes. Of course,
$\rho-\omega$ interference is included for \ee\ data only and the
corresponding amplitude ($\alpha_{\rho \omega}$) fitted. Fits are performed
separately for the $\tau$ and \ee\ form factors. The upper range of the fit
is taken at 3.6 GeV$^2$ for \ee, but at 2.4 GeV$^2$ for $\tau$ to avoid 
the extreme edge of the spectral function where large corrections are applied.

\begin{figure}[t]
   \centerline{\psfig{file=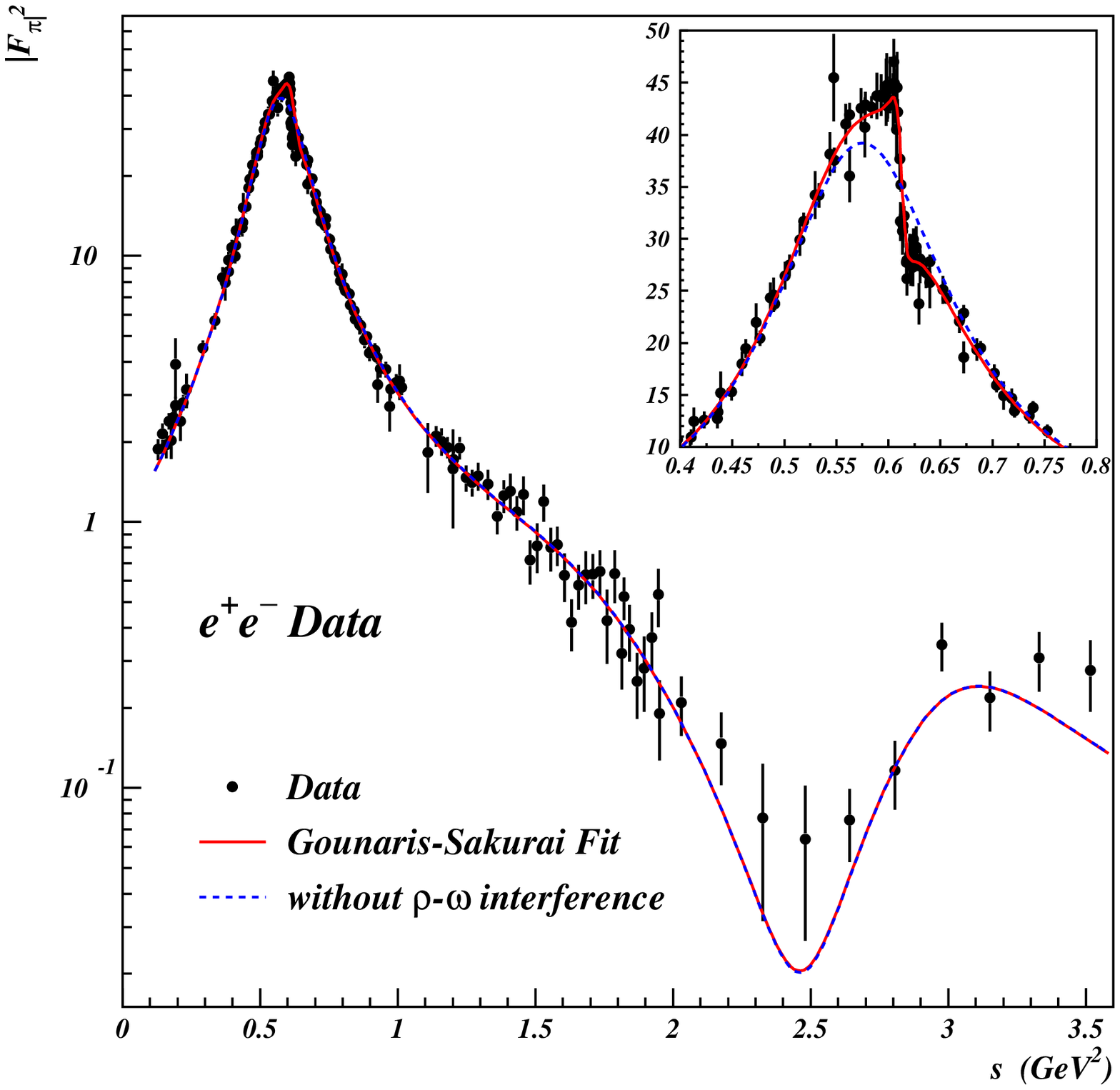,width=80mm}}
   \caption{The pion form factor squared $|F_\pi^0(s)|^2$ from \ee\ data 
   and the combined fit from Table~\ref{2pi_fits}.}
\label{fit_ee_2pi}
\end{figure}

\begin{figure}[t]
   \centerline{\psfig{file=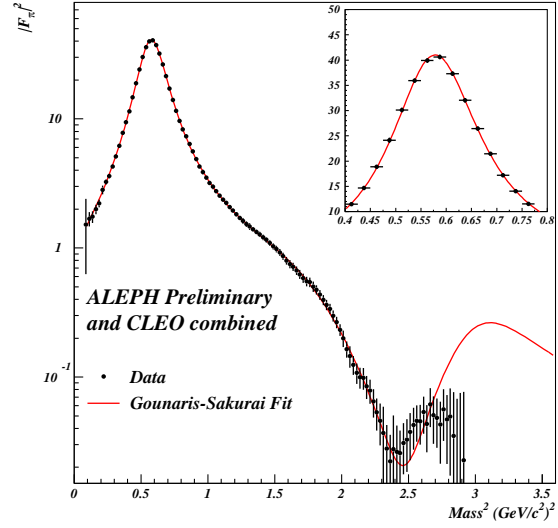,width=80mm}}
   \caption{The pion form factor squared $|F_\pi^-(s)|^2$ from $\tau$ data 
   and the combined fit from Table~\ref{2pi_fits}. }
\label{fit_tau_2pi}
\end{figure}

It turns out that the fitted $\rho$ masses and widths (essentially the latter)
are quite sensitive to the strength of the $\rho'$ and $\rho''$ amplitudes,
$\beta$ and $\gamma$. So, depending on the type of fit, the derived values
exhibit some systematic shifts. The \ee\ and $\tau$ fits yield significantly
different $\rho'$ amplitudes and phases: the phase in the $\tau$ fit is 
consistent with 180$^0$, while it comes out 30$^0$ smaller in the \ee\ fit.
This is merely a reflection of the discrepancy between the measurements
of the pion form factor, especially between 0.8 and 1.0~GeV.
A fit to both data sets is also performed, keeping common values for the
$\rho'$ and $\rho''$ parameters (second-order with respect to the dominant
investigated $\rho$ parameters).
Table~\ref{2pi_fits} presents the results of the common fits, the quality
of which can be inspected in Figs.~\ref{fit_ee_2pi} and \ref{fit_tau_2pi}. 

The differences found between the masses and widths of the charged and neutral
$\rho$'s are somewhat larger than predicted by Chiral Perturbation
Theory ($\chi$PT)~\cite{bijnens}($|m_{\rho^-}-m_{\rho^0}|<0.7$ MeV) and isospin
breaking ($\Gamma_{\rho^-}-\Gamma_{\rho^0}=(0.7\pm0.3)$ MeV). However,
if the mass difference (2.6~MeV) is taken as an experimental fact, 
then the larger observed width difference (3.1~MeV) could be explained by 
$\chi$PT (0.7 MeV from isospin breaking + 1.5 MeV from the mass difference).
Whether or not these discrepancies are of experimental or theoretical nature 
is still open~\cite{dehz,andreas_02} and needs to be further investigated.

\begin{table*}
\begin{center}
{
%\small
\begin{tabular}{|c|c|c|c|} \hline 
           &  $\tau$ & $e^+e^-$ & combined  \\
\hline \hline
$m_{\rho^0}$      & - & 773.3 $\pm$ 0.6 & 772.7 $\pm$ 0.5 \\
$\Gamma_{\rho^0}$ & - & 145.2 $\pm$ 1.3 & 146.4 $\pm$ 0.9 \\
$m_{\rho^-}$      & 775.0 $\pm$ 0.6 & - & 775.3 $\pm$ 0.6 \\
$\Gamma_{\rho^-}$ & 149.5 $\pm$ 1.1 & - & 149.5 $\pm$ 0.8 \\
\hline
$\alpha_{\rho\omega}$& - & $(2.02 \pm 0.10)~10^{-3}$ & $(1.98 \pm 0.10)~10^{-3}$ \\
\hline
$\beta$   & 0.195 $\pm$ 0.028 & 0.123 $\pm$ 0.011 & 0.172 $\pm$ 0.006 \\
$\phi_\beta$   & 173.0 $\pm$ 7.0 & 139.4 $\pm$ 6.5 & 178.2 $\pm$ 4.5 \\
$m_{\rho'}$   & 1440 $\pm$ 34 & 1337 $\pm$ 35 & 1415 $\pm$ 15 \\
$\Gamma_{\rho'}$  & 597 $\pm$ 102 & 569 $\pm$ 81 & 528 $\pm$ 42 \\   
\hline
$\gamma$   & 0.095 $\pm$ 0.029 & 0.048 $\pm$ 0.008 & 0.072 $\pm$ 0.006 \\
$\phi_\gamma$   & 0. & 0. & 0. \\
$m_{\rho''}$      & 1713 & 1713 $\pm$ 15 &  1741 $\pm$ 20 \\
$\Gamma_{\rho''}$   & 235 & 235 & 235 \\
\hline
\hline
\end{tabular}
}\caption{
         Results of fits to the pion form factor squared to $\tau$ and \ee\ 
         data (ALEPH and CLEO) separately, then combined. The parametrization 
         of the $\rho$ line shape (Gounaris-Sakurai) is described in the text. 
         All mass and width values are in MeV and the phase 
         $\phi_\beta$ in degrees.}
\label{2pi_fits}
\end{center}
\end{table*}

\subsection{The $4\pi$ Vector State}

The $4 \pi$ final states have also been studied~\cite{aleph_v,cleo_4pi}.
Tests of CVC are severely hampered by large deviations between different
$e^+e^-$ experiments which disagree beyond their quoted systematic
uncertainties. A CLEO analysis studies the resonant structure in the
$3 \pi \pi^0$ channel which is shown to be dominated by $\omega \pi$ and
$a_1 \pi$ contributions. The $\omega \pi$ spectral function  
is interpreted by
a sum of $\rho$-like amplitudes. The mass of the second state is however
found at $(1523 \pm 10)$MeV, in contrast with the value $(1406 \pm 14)$MeV
from the fit of the $2 \pi$ spectral function. This point has to be
clarified. Following a limit of $8.6\%$ obtained earlier by 
ALEPH~\cite{aleph_omega}, CLEO sets a new $95\%$ CL limit of $6.4\%$ for
the relative contribution of second-class currents in the decay
$\tau \rightarrow \nu_\tau \pi^- \omega$ from the hadronic angular decay
distribution.

\begin{figure}[t]
   \centerline{\psfig{file=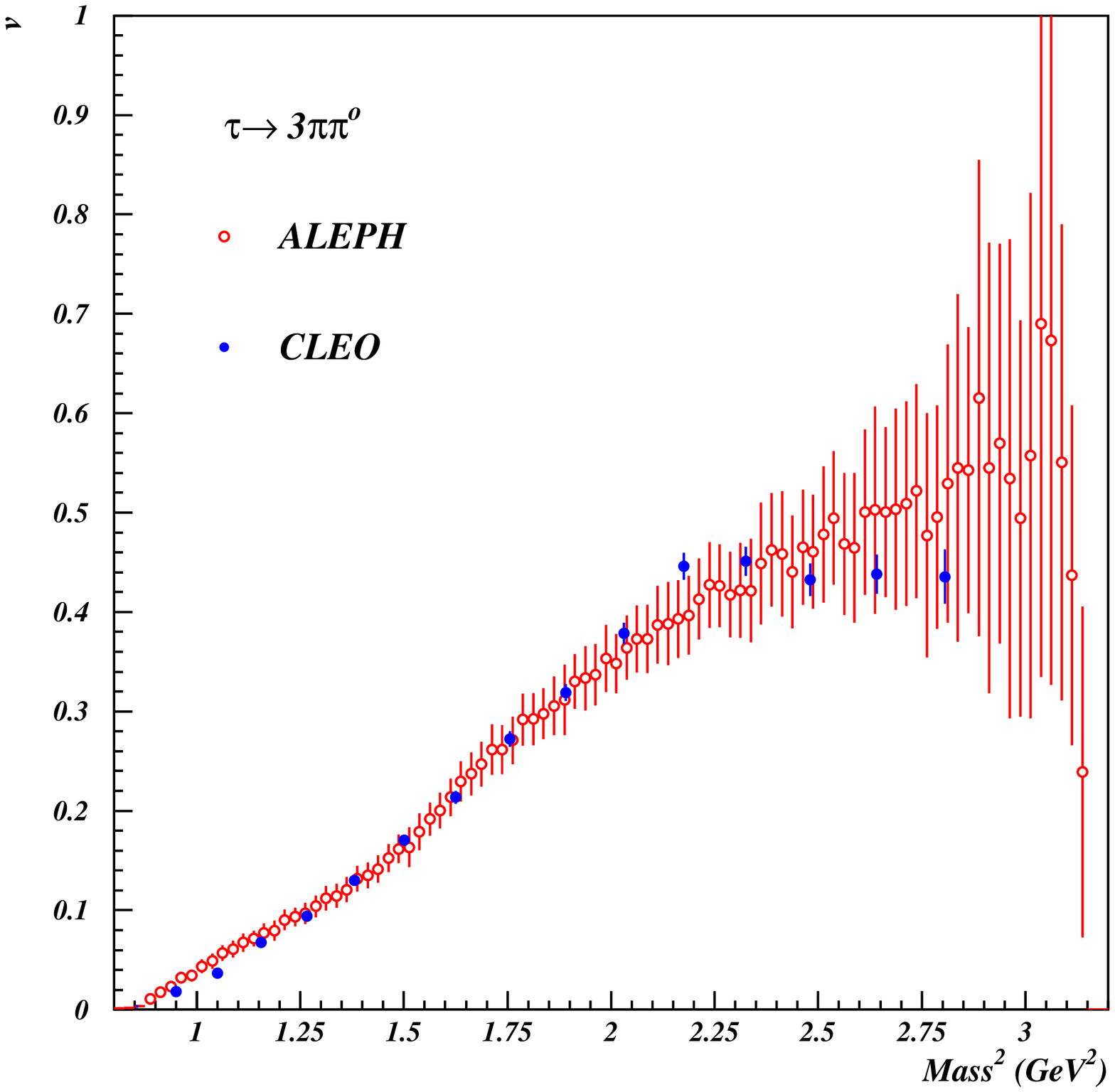,width=75mm}}
   \caption{The spectral function for the $3\pi \pi^0$ mode as
            determined by ALEPH and CLEO.}
\label{aleph_cleo_3pipi0}
\end{figure}

The new ALEPH spectral function in the $3\pi \pi^0$ mode is shown in 
Fig.~\ref{aleph_cleo_3pipi0}: it compares well with the CLEO published 
results~\cite{cleo_4pi}, except maybe in the threshold region near 1~GeV.

\begin{figure}[t]
   \centerline{\psfig{file=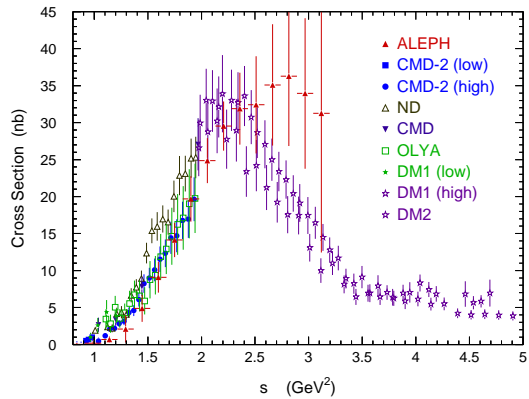,width=70mm}}
   \caption{The cross section for $e^+e^- \rightarrow 2\pi^+ 2\pi^-$ 
    from \ee\ experiments
    and derived from the $\pi 3\pi^0$ spectral function measured by ALEPH.}
\label{tau_ee_4pi}
\end{figure}

\begin{figure}[t]
   \centerline{\psfig{file=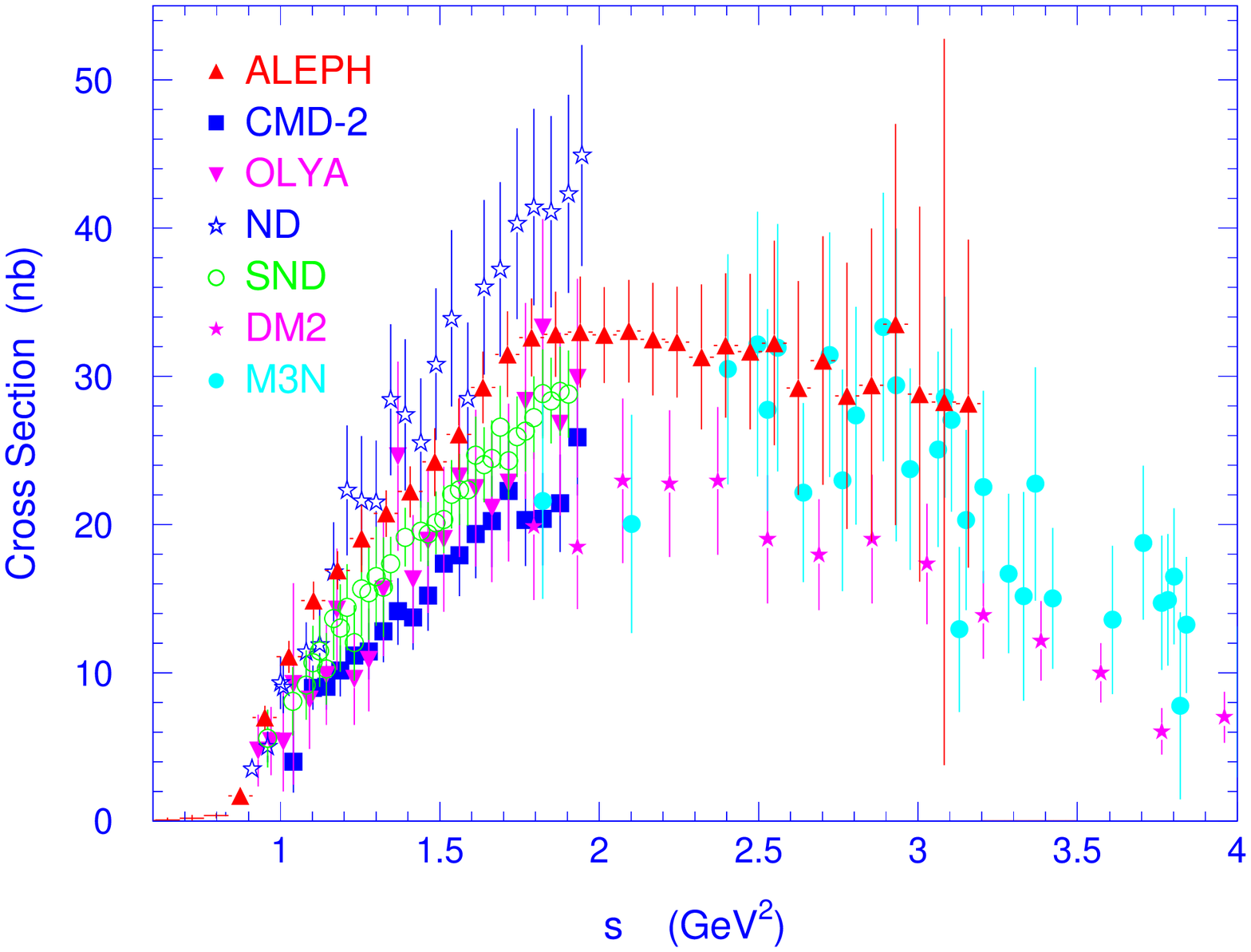,width=70mm}}
   \caption{The cross section for $e^+e^- \rightarrow \pi^+ \pi^- 2\pi^0$ 
    from \ee\ experiments and derived from the $\pi 3\pi^0$ 
    and $3\pi \pi^0$ spectral functions measured by ALEPH.}
\label{tau_ee_2pi2pi0}
\end{figure}

The $\tau$ spectral function in the four-pion modes can be used to predict
the corresponding isospin-rotated \ee\ cross sections using the SU(2)
symmetric relations
\beqns
 \sigma_{e^+e^-\rightarrow\,\pi^+\pi^-\pi^+\pi^-}^{I=1}
        & = &
             2\cdot\frac{4\pi\alpha^2}{s}\,
             v_{\pi^-\,3\pi^0}~, \\[0.3cm]
 \sigma_{e^+e^-\rightarrow\,\pi^+\pi^-\pi^0\pi^0}^{I=1}
        & = &
             \frac{4\pi\alpha^2}{s}\,
             \left[v_{2\pi^-\pi^+\pi^0}
                  -
                     v_{\pi^-\,3\pi^0}
             \right]
\eeqns
Comparisons can be performed after applying known isospin-breaking 
corrections~\cite{czyz,dehz}. As shown in Figs.~\ref{tau_ee_4pi} and
\ref{tau_ee_2pi2pi0}, agreement is observed in the $2\pi^+ 2\pi^-$
mode, whereas the situation is confused in the $\pi^+ \pi^- 2\pi^0$
mode where a large spread exists between the different \ee\ results
(references can be found in Ref.~\cite{dehz}).

\subsection{Axial-vector States}

The decay $\tau \rightarrow \nu_\tau 3\pi$ is the cleanest place to study
axial-vector resonance structure. The spectrum is dominated by the $1^+$ 
$a_1$ state, known to decay essentially through $\rho \pi$. A comprehensive
analysis of the $\pi^- 2\pi^0$ channel has been presented by CLEO. 
First, a model-independent determination of the hadronic structure 
functions gave no evidence for non-axial-vector contributions 
($<17\%$ at $90\%$ CL)~\cite{cleo_3pisf}. Second, a partial-wave
amplitude analysis was performed~\cite{cleo_3pi}: while the dominant 
$\rho \pi$ mode was of course confirmed, it came as a surprize that an 
important contribution ($\sim 20\%$) from scalars 
($\sigma$, $f_0(1470)$, $f_2(1270)$) was found in the $2\pi$ system.

The $a_1 \rightarrow \pi^- 2\pi^0$ precisely determined line shape  
shows the opening of the $K^*K$ decay mode in the total $a_1$ width.
The derived branching ratio, $B(a_1 \rightarrow K^*K)=(3.3 \pm 0.5)\%$
is in good agreement with ALEPH results on the $K \bar K \pi$ modes which
were indeed shown (with the help of $e^+e^-$ data and CVC) to be 
axial-vector ($a_1$) dominated with $B(a_1 \rightarrow K^*K)=(2.6 \pm 0.3)\%$
~\cite{aleph_ksum}. No conclusive evidence for a higher mass state ($a'_1$)
was found in this analysis.

\section{Inclusive Nonstrange Spectral Functions}

\begin{figure}
        \psfig{file=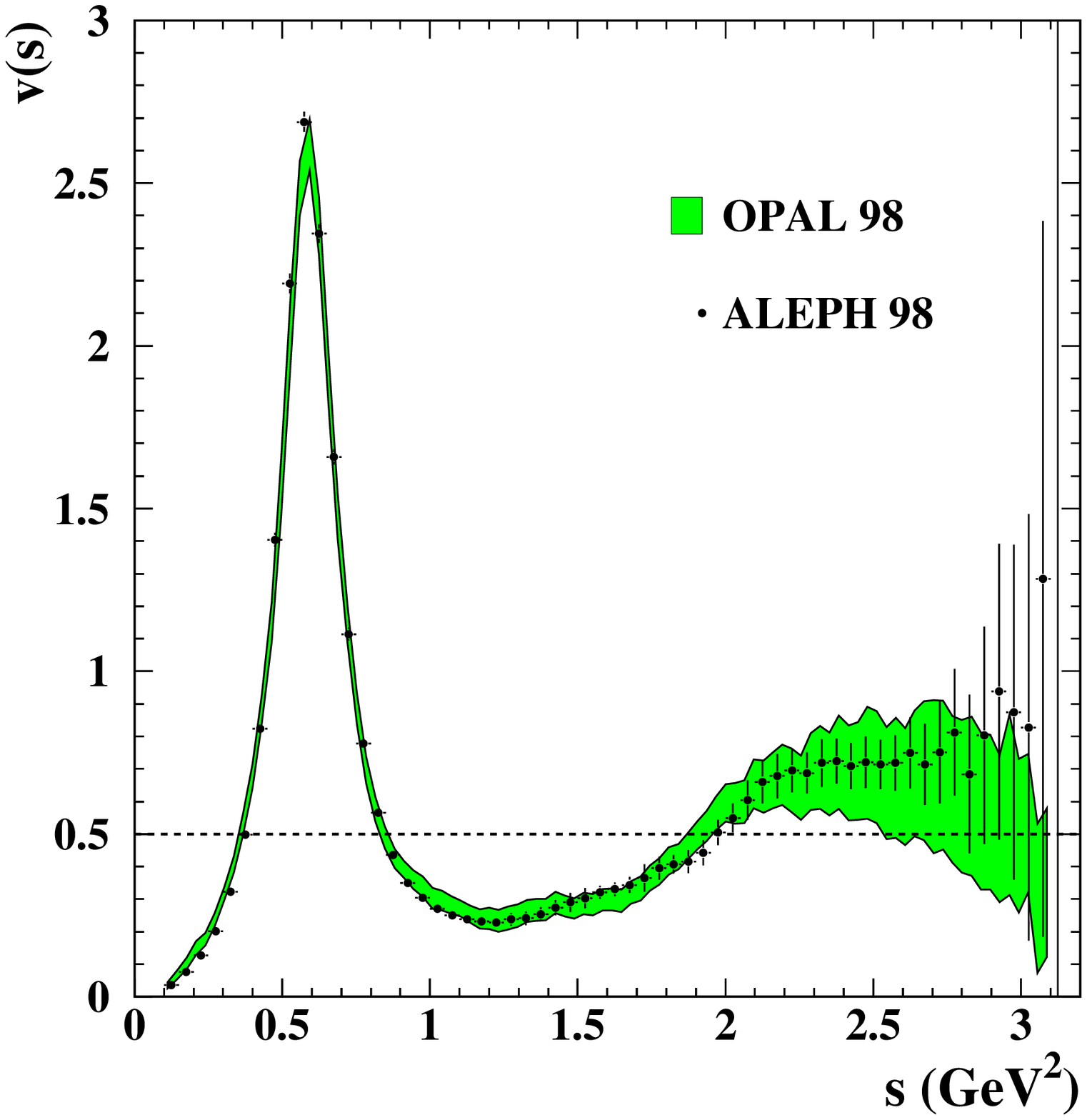,width=6cm}
%         bbllx=70,bblly=150,bburx=560,bbury=650}% 
%        \caption{Inclusive nonstrange vector spectral function from 
%         ALEPH and OPAL. The dashed line is the expectation from the naive 
%         parton model}%
%	\label{v_aleph_opal}
%\end{figure}

%\smallskip
%\begin{figure}
        \psfig{file=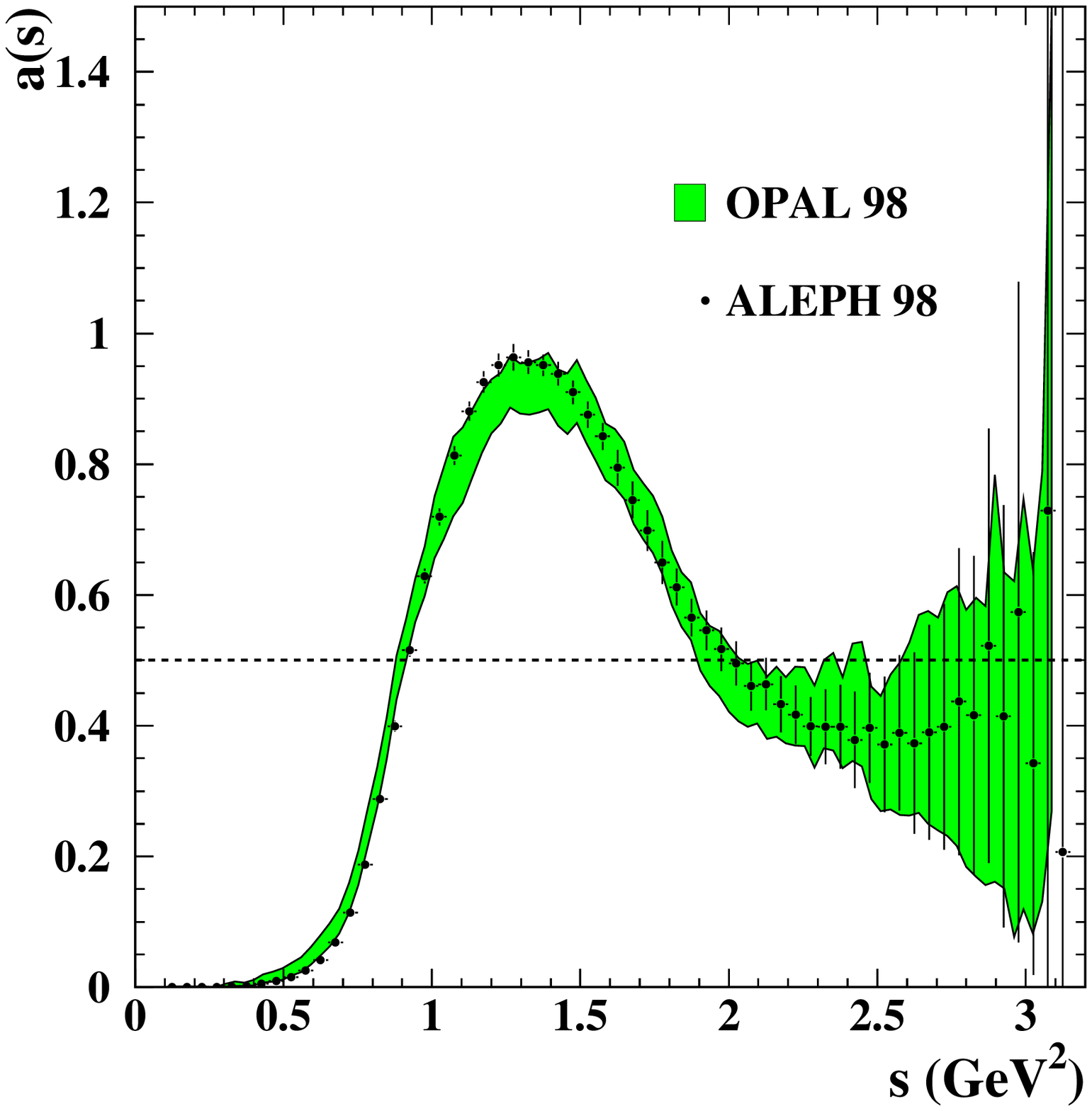,width=6cm}
%         bbllx=70,bblly=150,bburx=560,bbury=650}% 
        \caption{ Inclusive nonstrange vector (top) and
         axial-vector (bottom) spectral functions from 
         ALEPH and OPAL. The dashed line is the expectation from the naive 
         parton model.}%
	\label{aleph_opal}
\end{figure}

The $\tau$ nonstrange spectral functions have been measured by ALEPH
~\cite{aleph_v,aleph_a} and OPAL~\cite{opal}. The procedure requires a
careful separation of vector (V) and axial-vector (A) states involving
the reconstruction of multi-$\pi^0$ decays and the proper treatment of
final states with a $K \bar K$ pair. The $V$ and $A$ spectral 
functions are given in Fig.~\ref{aleph_opal}.
They show a strong resonant behaviour, dominated by the lowest 
$\rho$ and $a_1$
states, with a tendancy to converge at large mass toward a value near
the parton model expectation. Yet, the vector part stays clearly above
while the axial-vector one lies below. Thus, the two spectral functions
are clearly not 'asymptotic' at the $\tau$ mass scale.

\smallskip

\begin{figure}[t]
   \centerline{\psfig{file=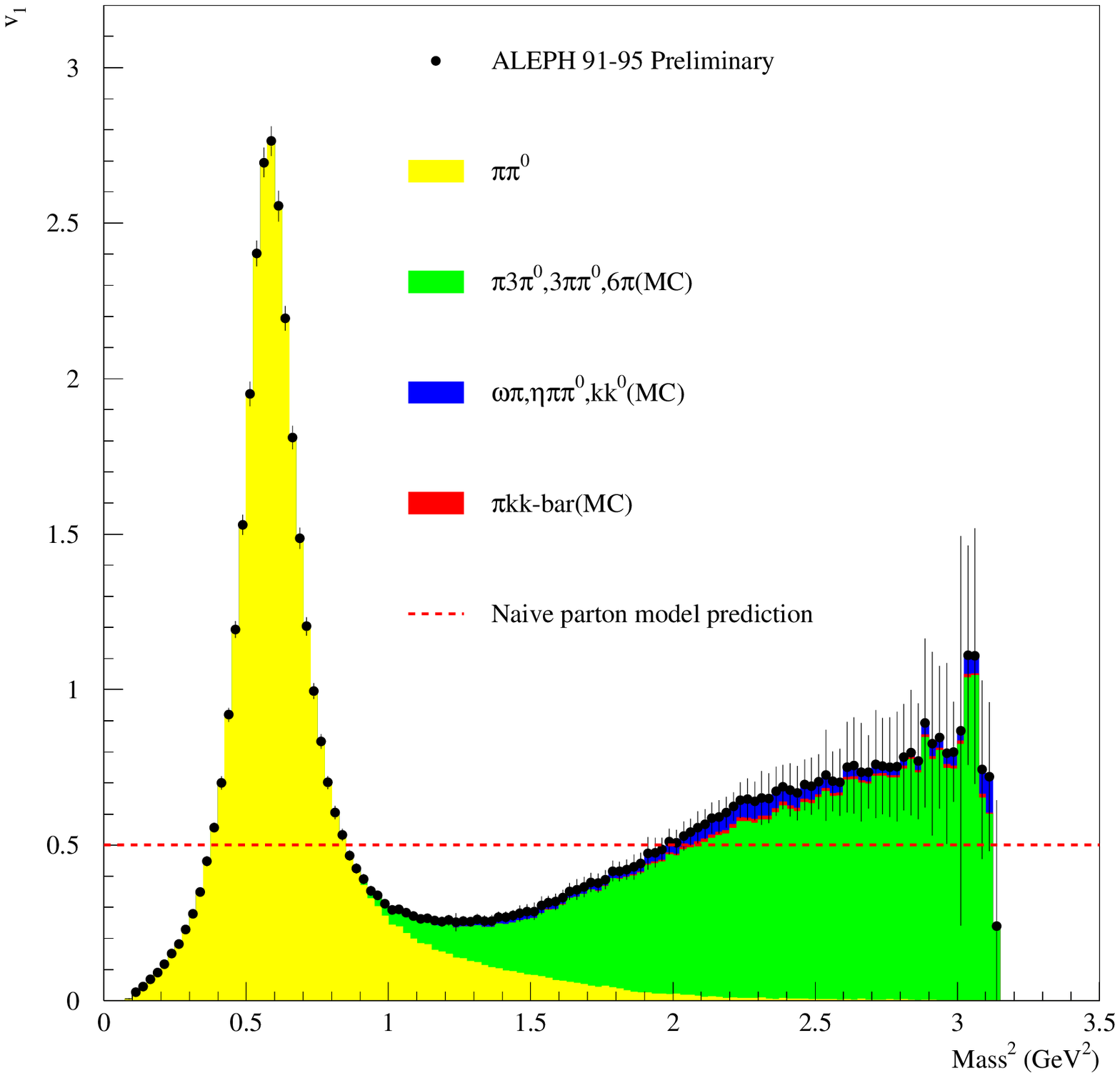,width=70mm}}
   \caption{Preliminary ALEPH inclusive vector spectral function with its
    different contributions. The dashed line is the expectation from 
    the naive parton model.}
\label{v_aleph}
\end{figure}

\begin{figure}[t]
   \centerline{\psfig{file=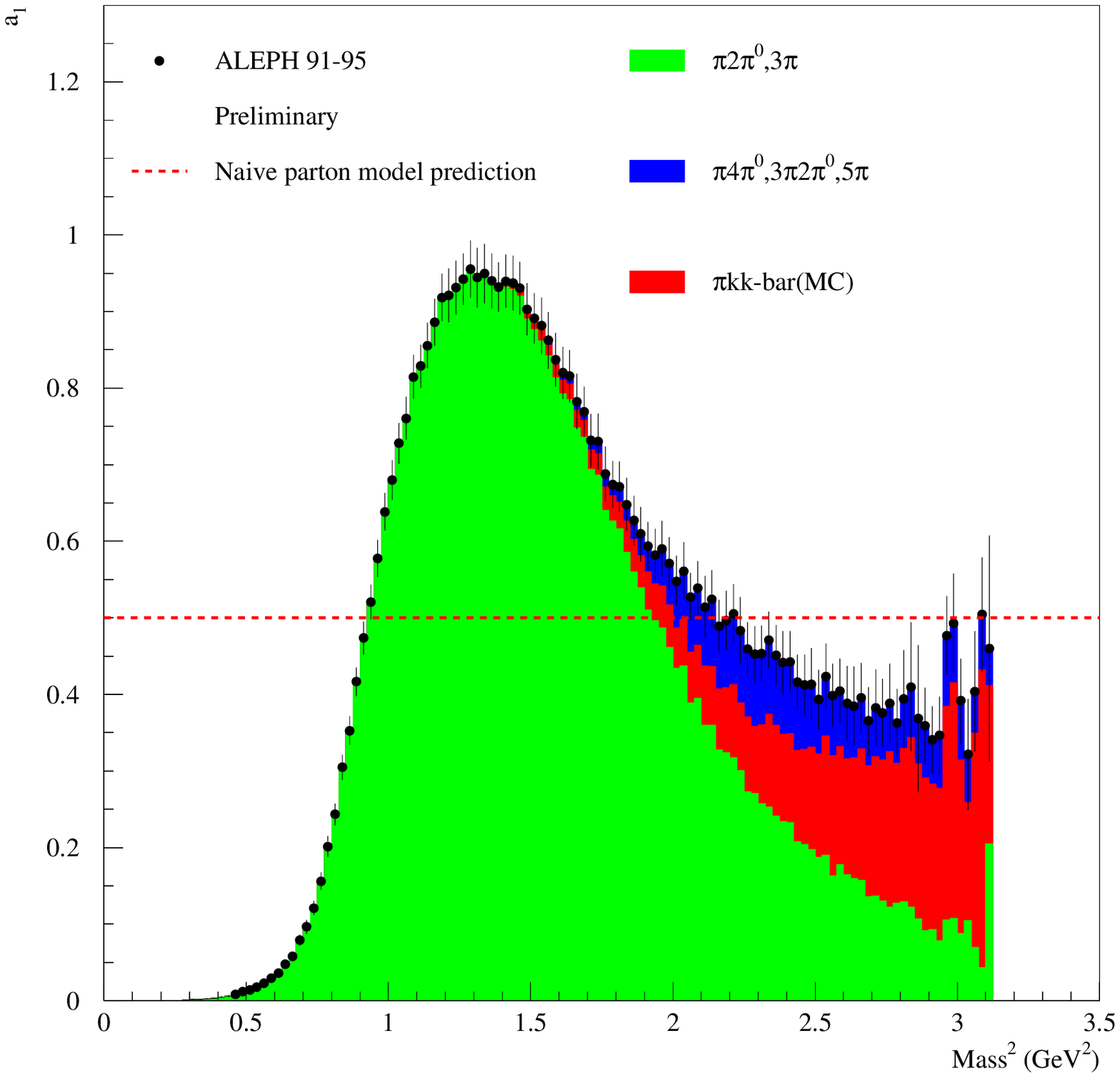,width=70mm}}
   \caption{Preliminary ALEPH inclusive axial-vector spectral function with its
    different contributions. The dashed line is the expectation from 
    the naive parton model. }
\label{a_aleph}
\end{figure}

The new inclusive ALEPH $V$ and $A$ spectral functions are given in 
Figs.~\ref{v_aleph} and \ref{a_aleph} with a breakdown of the respective
contributions.

\begin{figure}[t]
   \centerline{\psfig{file=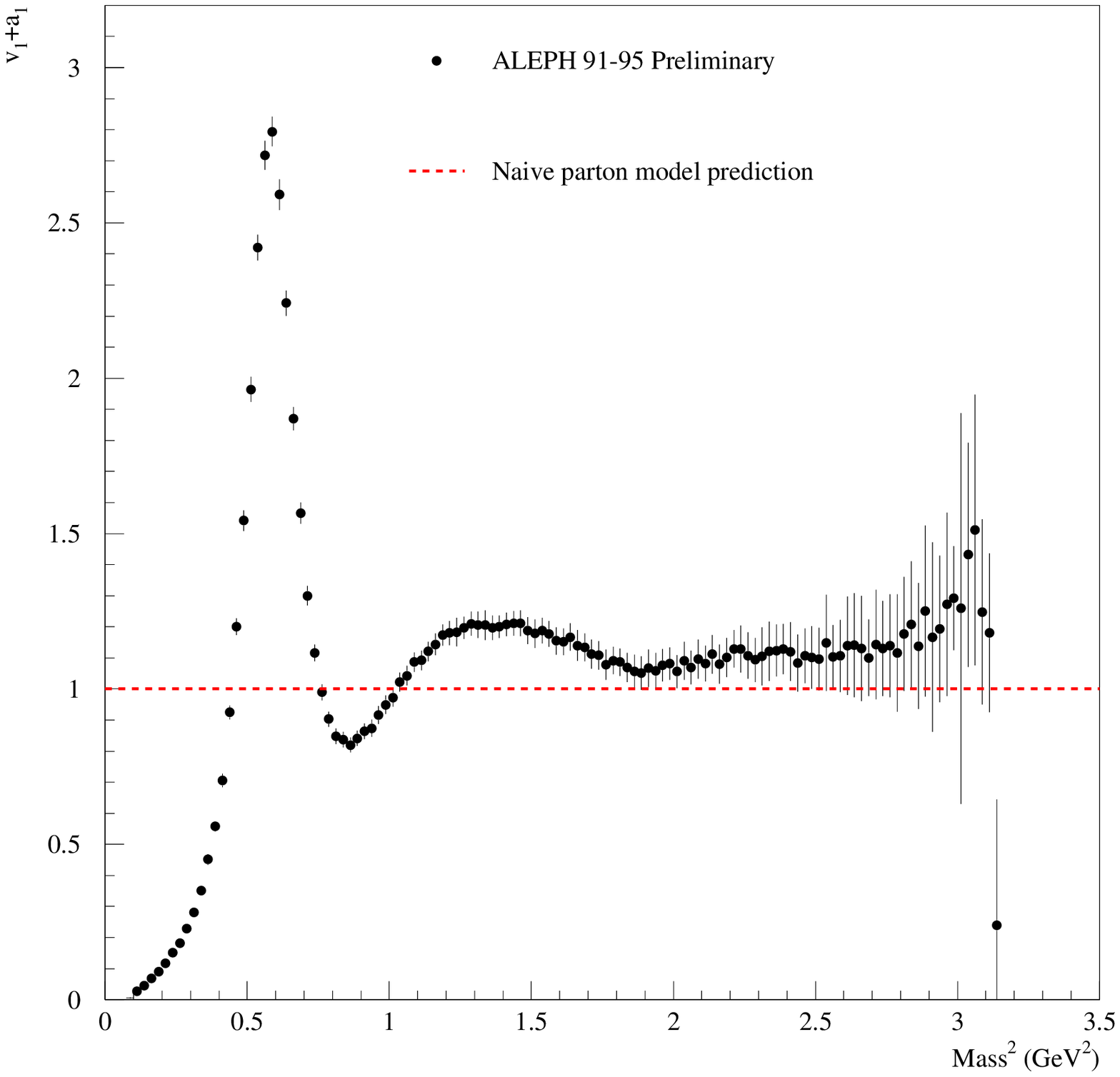,width=70mm}}
   \caption{The preliminary inclusive $V+A$ nonstrange spectral function 
      from ALEPH. The dashed line is the expectation from the naive 
      parton model.}
\label{vpa_aleph}
\end{figure}

The $V+A$ spectral function, shown in Fig.~\ref{vpa_aleph} has a clear 
pattern converging toward a value above the parton level as expected in
QCD. In fact, it displays a textbook example of global duality, since
the resonance-dominated low-mass region shows an oscillatory behaviour around
the asymptotic QCD expectation, assumed to be valid in a local sense only
for large masses. This feature will be quantitatively discussed in the 
next section.

\section{QCD Analysis of Nonstrange $\tau$ Decays}

\subsection{Motivation}

The total hadronic $\tau$ width, properly normalized to the known leptonic
width,

\beq
     R_\tau = \frac{\Gamma(\tau^-\rightarrow{\rm hadrons}^-\,\nu_\tau)}
                   {\Gamma(\tau^-\rightarrow e^-\,\bar{\nu}_e\nu_\tau)}
\eeq
should be well predicted by QCD as it is an inclusive observable. Compared
to the similar quantity defined in \ee\ annihilation, it is even twice
inclusive: not only are all produced hadronic states at a given mass summed
over, but an integration is performed over all the possible masses from
$m_{\pi}$ to $m_{\tau}$.

This favourable situation could be spoiled by the fact that the $Q^2$ scale
is rather small, so that questions about the validity of a perturbative
approach can be raised. At least two levels are to be considered: the 
convergence of the perturbative expansion and the control of the
nonperturbative contributions. Happy circumstances make
these latter components indeed very small~\cite{braaten88,narpic88}.

\subsection{Theoretical Prediction for \boldmath\Rt}

The imaginary parts of the vector 
and axial-vector two-point correlation functions 
$\Pi^{(J)}_{\bar{u}d,V/A}(s)$, with the spin $J$ of the hadronic 
system, are proportional to the $\tau$ hadronic \sfs\ with 
corresponding quantum numbers. The non-strange ratio \Rt\
can be written as an integral of these \sfs\ over the 
invariant mass-squared $s$ of the f\/inal state hadrons~\cite{bnp}:
\begin{eqnarray}
\label{eq_rtauth1}
   R_\tau(s_0) &=&
     12\pi S_{\rm EW}\intl_0^{s_0}\frac{ds}{s_0}\left(1-\frac{s}{s_0}
                                    \right)^{\!\!2} \\
   & & \hspace{-2.3cm} 
     \times \left[\left(1+2\frac{s}{s_0}\right){\rm Im}\Pi^{(1)}(s+i\e)
      \,+\,{\rm Im}\Pi^{(0)}(s+i\e)\right] \nonumber
\end{eqnarray}
By Cauchy's theorem the imaginary part of $\Pi^{(J)}$ is 
proportional to the discontinuity across the positive real axis. 

The energy scale $s_0= m_\tau^2$ is large enough so that
contributions from nonperturbative ef\/fects are small. It is therefore 
assumed that one can use the {\it Operator Product Expansion} (OPE) 
to organize perturbative and nonperturbative contributions~\cite{svz} to 
\Rts.

The theoretical prediction of the vector and axial-vector
ratio \RtVA\ can thus be written as:
\begin{eqnarray}
\label{eq_delta}
   R_{\tau,V/A} \;=\;
     \frac{3}{2}|V_{ud}|^2S_{\rm EW} \\
   & & \hspace{-4.1cm}
     \times \left(1 + \delta^{(0)} + 
     \delta^\prime_{\rm EW} + \delta^{(2-\rm mass)}_{ud,V/A} + 
     \hm\hm\sum_{D=4,6,8}\hm\hm\hm\hm\delta_{ud,V/A}^{(D)}\right) \nonumber
\end{eqnarray}
with the residual non-logarithmic electroweak
correction $\delta^\prime_{\rm EW}=0.0010$~\cite{braaten}, 
neglected in the following, and the dimension $D=2$ 
contribution $\delta^{(2-\rm mass)}_{ud,V/A}$ 
from quark masses which is lower than $0.1\%$ for $u,d$ quarks.
The term $\delta^{(0)}$ is the purely perturbative 
contribution, while the $\delta^{(D)}$ are the OPE
terms in powers of $s_0^{-D/2}$ of the following form
\beq
\label{eq_ope}
    \delta_{ud,V/A}^{(D)} \;\sim\;
       \hm\hm\hm\sum_{{\rm dim}{\cal O}=D}\hm\hm\hm
            \frac{\langle{\cal O}_{ud}\rangle_{V/A}}
                 {(-s_0)^{D/2}}
\eeq
where the long-distance nonperturbative ef\/fects are absorbed into
the vacuum expectation elements $\langle{\cal O}_{ud}\rangle$.

The perturbative expansion (FOPT) is known to third order~\cite{3loop}.
A resummation of all known higher order 
logarithmic integrals improves the convergence 
of the perturbative series (contour-improved method \FOPTCI)~\cite{pert}. 
As some ambiguity persists, the results are given as an average of the
two methods with the difference taken as a systematic uncertainty.

\subsection{Measurements}

The QCD analysis of the $\tau$ hadronic width has not yet been completed 
with the final ALEPH spectral functions. Results given below correspond
to the published analyses with a smaller data set.

The ratio \Rt\ is obtained from measurements of the leptonic branching 
ratios:
\beqn
  R_\tau&=&3.647\pm0.014 
\eeqn
using a value 
$B(\tau^-\rightarrow e^-\,\bar{\nu}_e\nu_\tau)=(17.794\pm0.045)\%$
which includes the improvement in accuracy provided by the 
universality assumption of leptonic currents together with the 
measurements of $B(\tau^-\rightarrow e^-\,\bar{\nu}_e\nu_\tau)$,
$B(\tau^-\rightarrow \mu^-\,\bar{\nu}_\mu\nu_\tau)$ and the $\tau$ 
lifetime. The nonstrange part of \Rt\ is obtained by subtracting
out the measured strange contribution (see last section).

Two complete analyses of the $V$ and $A$ parts have been performed by
ALEPH~\cite{aleph_a} and OPAL~\cite{opal}. Both use the world-average
leptonic branching ratios, but their own measured spectral functions.
The results on $\alpha_s(m_\tau^2)$ are therefore strongly correlated
and indeed agree when the same theoretical prescriptions are used. 

\subsection{Results of the Fits}

The results of the fits are given in Table~\ref{tab_asresults} for
the ALEPH analysis. Similar results are obtained by OPAL. It is worth
emphasizing that the nonperturbative contributions are found to be
very small, as expected. The limited number of observables
and the strong correlations between the spectral moments introduce
large correlations, especially between the f\/itted 
nonperturbative operators.

\begin{table}
{\small
  \begin{tabular}{|l||c|c|} \hline 
     & &  \\
 ALEPH&$\alpha_s(m_{\tau}^2)$       &$\delta_{\rm NP}$\\
\hline 
 V    &  $0.330\pm0.014\pm0.018$    &  $0.020\pm0.004$  \\
 A    &  $0.339\pm0.013\pm0.018$    &  $-0.027\pm0.004$  \\
\hline
 V+A  &  $0.334\pm0.007\pm0.021$    &  $-0.003\pm0.004$  \\
\hline
  \end{tabular}
}
  \caption{
              F\/it results of \asm\ and the OPE nonperturbative 
              contributions from vector, axial-vector and $(V+A)$ combined
              fits using the corresponding ratios \Rt\ and the spectral 
              moments as input parameters. The second error is given for 
              theoretical uncertainty.}
\label{tab_asresults}
\end{table}

One notices a remarkable agreement within statistical errors
between the \asm\ values using vector and axial-vector data.
The total nonperturbative power contribution to \RtVpA\ is compatible 
with zero within an uncertainty of 0.4\pc, that is much smaller than 
the error arising from the perturbative term. This cancellation of the 
nonperturbative terms increases the confidence on the \asm\ 
determination from the inclusive $(V+A)$ observables.

The f\/inal result from ALEPH is : 
\beq
\label{eq_asres}
   \alpha_s(m_\tau^2)    = 0.334 \pm 0.007_{\rm exp} 
                                 \pm 0.021_{\rm th} 
\eeq
where the f\/irst error accounts for the experimental uncertainty and 
the second gives the uncertainty of the theoretical prediction of
$R_\tau$ and the spectral moments as well as the ambiguity of the 
theoretical approaches employed.

In the OPAL analysis the corresponding results are quoted within three 
prescriptions for the perturbative expansion, respectively
\FOPTCI, FOPT and with renormalon chains, {\it i.e.}
\begin{eqnarray}
\label{qcd_opal}
\alpha_s(m_\tau^2)&=&0.348\pm 0.009_{\rm exp}\pm 0.019_{\rm th} \\
                  &=&0.324\pm 0.006_{\rm exp}\pm 0.013_{\rm th} \\
                  &=&0.306\pm 0.005_{\rm exp}\pm 0.011_{\rm th}
\end{eqnarray}

\subsection{Test of the Running of \boldmath$\alpha_s(s)$ at Low Energies}

Using the \sfs, one can simulate the physics of a hypothetical 
$\tau$ lepton with a mass $\sqrt{s_0}$ smaller than $m_\tau$
through equation~(\ref{eq_rtauth1}) and hence further investigate QCD
phenomena at low energies. Assuming quark-hadron duality, 
the evolution of $R_\tau(s_0)$ provides a direct test of the 
running of $\alpha_s(s_0)$, governed by the RGE $\beta$-function. 
On the other hand, it is a test of the validity of the OPE approach 
in $\tau$ decays. 

\smallskip
\begin{figure}
        \psfig{file=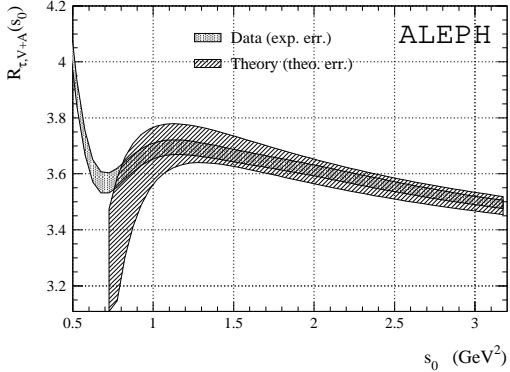,width=7cm}% 
        \caption{
              The ratio \RtVpA\ versus the square ``$\tau$ mass'' $s_0$. 
              The curves are plotted as error bands to emphasize their 
              strong point-to-point correlations in $s_0$. Also 
              shown is the theoretical prediction using 
              the results of the fit at $s_0=m_\tau^2$.}%
	\label{fig_rtau}
\end{figure}

\smallskip
\begin{figure}
        \psfig{file=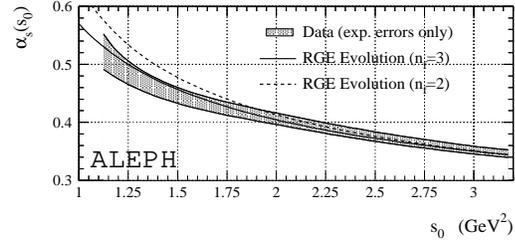,width=7cm}% 
        \caption{
              The running of $\alpha_s(s_0)$ obtained from the 
              fit of the theoretical prediction to \RtVpAs.
              The shaded band shows the data including experimental
              errors. The curves give the four-loop RGE evolution 
              for two and three flavours.}%
	\label{fig_runas}
\end{figure}

The functional dependence of \RtVpAs\ is plotted in 
Fig.~\ref{fig_rtau} together with the theoretical 
prediction using the results of Table~\ref{tab_asresults}.
Below $1~{\rm GeV}^2$ the error of the theoretical 
prediction of \RtVpAs\ starts to blow up due to the 
increasing uncertainty from the unknown fourth-order perturbative term.
Fig.~\ref{fig_runas} has the same physical content as 
Fig.~\ref{fig_rtau}, but translated into the running 
of $\alpha_s(s_0)$, \ie, the experimental value for $\alpha_s(s_0)$ 
has been individually determined at every $s_0$ from the comparison 
of data and theory. Good agreement is observed with the four-loop 
RGE evolution using three quark f\/lavours.

The experimental fact that the nonperturbative contributions 
cancel over the whole range $1.2~{\rm GeV}^2\le s_0\le m_\tau^2$ 
leads to conf\/idence that the \as\ determination from the inclusive 
$(V+A)$ data is robust.  

\subsection{Discussion on the Determination of $\alpha_s(m_\tau^2)$}

The evolution of the 
\asm\ measurement from the inclusive $(V+A)$ observables based on 
the Runge-Kutta integration of the dif\/ferential equation 
of the renormalization group to 
N$^3$LO~\cite{alpha_evol,pichsanta} yields for the ALEPH analysis
\beq
\label{alphaevol}  
   \alpha_s(M_{\rm Z}^2) = \nonumber \\
                           0.1202 \pm 0.0008_{\rm exp} 
                                  \pm 0.0024_{\rm th} 
                                  \pm 0.0010_{\rm evol}
\eeq
where the last error stands for
possible ambiguities in the evolution due to uncertainties in the 
matching scales of the quark thresholds~\cite{pichsanta}. 

\smallskip
\begin{figure}
        \psfig{file=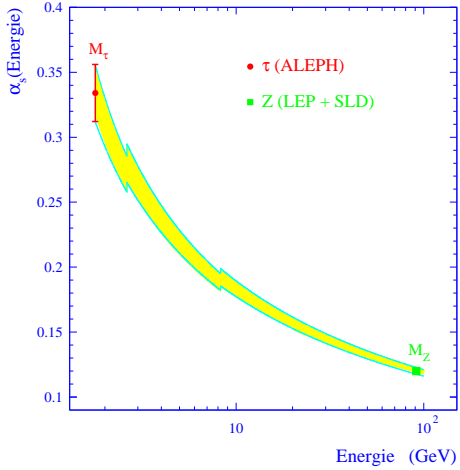,width=6cm}% 
        \caption{
     Evolution of the strong coupling (measured at $m_\tau^2)$
     to $M_Z^2$ predicted by QCD compared to the direct measurement.
     The evolution is carried out at 4 loops, while the flavour
     matching is accomplished at 3 loops at $2~m_c$ and $2~m_b$
     thresholds.}%
	\label{alphatz}
\end{figure}

The result (\ref{alphaevol}) can be compared to the precise determination 
from the measurement of the $Z$ width, as obtained in the global 
electroweak fit. The variable $R_Z$ has similar advantages to $R_\tau$,
but it differs concerning the convergence of the perturbative
expansion because of the much larger scale. It turns out that this
determination is dominated by experimental errors with very small
theoretical uncertainties, \ie\ the reverse of the situation 
encountered in $\tau$ decays.
The most recent value~\cite{ewfit} yields 
$\alpha_s(M_{\rm Z}^2) = 0.1183 \pm 0.0027$, in excellent agreement
with (\ref{alphaevol}). 
Fig.~\ref{alphatz} illustrates well the agreement between the evolution
of $\alpha_s(m_{\tau}^2)$ predicted by QCD and
$\alpha_s(M_{\rm Z}^2)$. 

\section{Strange Spectral Function and Strange Quark Mass}

\begin{figure}
        \psfig{file=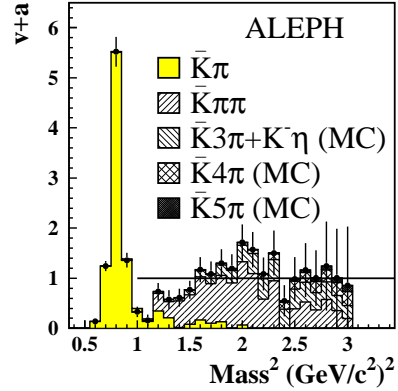,width=6.5cm}% 
        \caption{The ALEPH strange spectral function with various 
        contributions indicated. Higher-mass contributions from final states 
        with larger multiplicities lead to very small branching fractions and 
        are estimated.}%
	\label{strange_sf}
\end{figure}

The spectral function for strange final states has been determined by
ALEPH~\cite{aleph_ksum}. Fig.~\ref{strange_sf} shows that it is 
dominated by the vector $K^*(890)$ and higher mass (mostly axial-vector)
resonances, leading into a poorly defined continuum region, however in 
agreement with the quark model.

\smallskip

The total rate for strange final states, using the complete ALEPH analyses 
supplemented by results from other experiments~\cite{dchpp} is determined to 
be $B(\tau \rightarrow \nu_\tau {\rm hadrons}_{S=-1})=(29.3\pm1.0)~10^{-3}$,
leading to
\beq
 R_{\tau,S}= 0.163 \pm 0.006.
\eeq

\begin{figure}
        \psfig{file=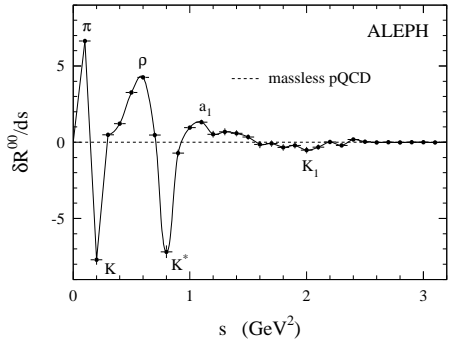,width=6cm}% 
        \caption{Differential rate for $\Delta_\tau^{00}$, difference between
       properly normalized nonstrange and strange spectral functions (see text
       for details). The contribution from massless perturbative QCD vanishes.
       To guide the eye, the solid line interpolates between bins of constant
       0.1 GeV$^2$ width.}%
	\label{diff_strange}
\end{figure}

Spectral moments are again useful tools to unravel the different components
of the inclusive rate. Since we are mostly interested in the specific
contributions from the $\overline{u}s$ strange final state, it is useful
to form the difference
\beq
 \Delta_\tau^{kl} \;\equiv\;
     \frac{1}{|V_{ud}|^2}R_{\tau,S=0}^{kl} - 
     \frac{1}{|V_{us}|^2}R_{\tau,S=-1}^{kl}
\eeq
where the flavour-independent perturbative part and gluon condensate cancel.
Fig.~\ref{diff_strange} shows the interesting behaviour of $\Delta_\tau^{00}$
expressed differentially as a function of $s$.
\smallskip

The leading QCD contribution to $\Delta_\tau^{kl}$ is a term proportional
to the square of the strange quark mass at the $\tau$ energy scale. Special
attention must be devoted to the perturbative expansion of this term which
has the early behaviour of an asymptotic series. We quote here the recent
result from the analysis of Ref.~\cite{cdghpp}, yielding
\beq
  m_s(m_\tau^2)~=~
       (120 \pm 11_{\rm exp} \pm 8_{V_{us}} \pm 19_{th})~{\rm MeV}\\
\eeq
where the dominant uncertainty is from theory, mostly because of the poor
convergence behaviour. Stability checks have been performed and their effect
has been conservatively taken into account in the estimate of the theoretical
uncertainty.

\section{Conclusions}

The decays $\tau \rightarrow \nu_\tau$ + hadrons constitute a clean and
powerful way to study hadronic physics up to $\sqrt{s} \sim 1.8$ GeV.
Beautiful resonance analyses have already been done, providing new
insight into hadron dynamics. Probably the major surprise has been the 
fact that inclusive hadron production is well described by perturbative
QCD with very small nonperturbative components at the $\tau$ mass.
Despite the fact that this low-energy region is dominated by resonance
physics, methods based on global quark-hadron duality work indeed very
well.

The new ALEPH preliminary results using the full LEP1 sample have been 
presented. Satisfactory agreement with CLEO is observed in the $\pi \pi^0$
and $3\pi \pi^0$ decay modes. The $\tau$ spectral functions
have now reached a precision level where detailed investigations are
possible, particularly in the most interesting $\pi \pi^0$ channel. The
breaking of SU(2) symmetry can be directly determined through the 
comparison between \ee\ and $\tau$ spectral functions. The difference
between the two spectral functions does not agree with the computed effect 
of isospin symmetry breaking and requires further experimentation.
The determination of the widths of the charged and the neutral $\rho$
is quite sensitive to the fit conditions, even though the data cover
a very large mass range.

The measurement of the vector and axial-vector spectral functions has
provided the way for quantitative QCD analyses. These spectral functions
are very well described in a global way by $O(\alpha_s^3)$ perturbative
QCD with small nonperturbative components. Precise determinations of
$\alpha_s$ agree for both spectral functions and they also agree with all
the other determinations from the Z width, the rate of Z to jets and
deep inelastic lepton scattering. Indeed from $\tau$ decays
\beq
      \alpha_s(M_{\rm Z}^2)_\tau=0.1202 \pm 0.0027
\eeq
in excellent agreement with the average from all other 
determinations~\cite{alphamor}
\beq
      \alpha_s(M_{\rm Z}^2)_{non-\tau}=0.1187 \pm 0.0020
\eeq

The strange spectral function yields a very competitive value for the
strange quark mass which can be evolved to the usual comparison scale
\beq
   m_s(2~{\rm GeV})=(116^{+20}_{-25})~{\rm MeV}
\eeq

The use of the vector $\tau$ spectral function and the QCD-based approach
as tested in $\tau$ decays improve the calculations of hadronic vacuum
polarization considerably. Significant results have been obtained for
the running of $\alpha$ to the Z mass and the muon anomalous magnetic
moment. Both of these quantities must be known with high precision as
they have the potential to give access to new physics.

\section*{Acknowledgments}

I would like to thank my colleagues Shaomin Chen, Andreas H\"ocker,
Changzheng Yuan and Zhiqing Zhang for their many contributions 
to this work. Fruitful discussions 
with V.~Cirigliano, G.~Ecker, S.I.~Eidelman, W. Marciano, H.~Neufeld, 
A.~Pich and A.~Weinstein are acknowledged. Congratulations to Abe Seiden 
and his staff for organizing an exciting Tau2002 Workshop.

\end{document}